\documentclass[a4paper]{jpconf}
\def \als {\alpha_{\mathrm{s}}}
\usepackage{graphicx}
\begin{document}
\title{The relation between cross-section, decay width and imaginary potential of heavy quarkonium in a quark-gluon plasma}
\author{Miguel \'{A}ngel Escobedo}
\address{Physik-Department, Technische Universit\"{a}t M\"{u}nchen, James-Franck-Str. 1, 85748 Garching, Germany}
\ead{miguel.escobedo@ph.tum.de}
\begin{abstract}
Heavy quarkonium suppression was proposed long ago as a signal of the formation of a deconfined phase in heavy ion collisions. Originally the mechanism responsible for this suppression was thought to be color screening. However perturbative computations and recent lattice studies suggest the existence of an imaginary part of the potential which could have a more important role for suppression than screening. In this work we review some general aspects of effective field theories for heavy quarkonium in the medium and discuss the physical phenomena behind the imaginary part of the potential and the decay width of heavy quarkonium and their corresponding cross sections. 
\end{abstract}
\section{Introduction}
Heavy quarkonium is a meson formed by two heavy quarks whose mass $m_Q$ is much bigger than $\Lambda_{QCD}$. They are quite different to other mesons. The first difference is that a lot of the physics relevant to study this system happens at a energy scale $m_Q$ where perturbation theory is applicable. For example the typical size of the more deeply bounded quarkonium states is of order $1/(m_Q v)$, where $v$ is the velocity of the heavy quarks around the center of mass, while for other mesons the size is of order $1/\Lambda_{QCD}$. A second difference is that heavy quarkonium is a non-relativistic system, meaning that $v\ll 1$. For this reason it is expectable that at some accuracy it can be described by a Schr\"{o}dinger equation with some potential, hence we can say that it is the QCD analogous to the hydrogen atom. On the other hand this small $v$ can spoil naive perturbation theory in a way that will be discussed later.

The original idea of using quarkonium suppression as a probe of deconfinement in heavy ion collisions is found in \cite{Matsui:1986dk}.  Since then this phenomena has been studied experimentally in many facilities, as for example SPS, RHIC and LHC. In the future the CBM experiment will be able to probe the situation when the quark-gluon plasma has a large chemical potential. Even though quarkonium suppression is a well established experimental fact an understanding of the responsible mechanism and a quantitative description of experimental data is still missing. 

The dissociation mechanism that was originally considered in \cite{Matsui:1986dk} was color screening. In order to illustrate this mechanism we can make use of what happens in the perturbative limit of QCD. There the potential between two infinitely heavy color charges in the vacuum will have the form of a Coulomb potential $V(r)=-\frac{\alpha}{r}$. Then the two heavy charges will feel an attractive force. If instead of the vacuum we are in a medium in which color screening takes place then we will have instead of a Coulomb a Yukawa potential $V(r)=-\frac{\alpha}{r}e^{-m_D r}$ where $m_D$ is a Debye mass that will encode the information about the medium. If the two charges are separated by a distance that is much bigger than $1/m_D$ they are not going to see each other so in an effective way they will behave like individual particles. 

However another mechanism for dissociation might also exists. The heavy quarkonium state will have some probability to be hit by some of the particles in the medium and this hit might induce a decay. This means that the quarkonium state might still exist but with a very short mean life. Indeed in a perturbative computation in \cite{Laine:2006ns} it was shown that the correlator of the heavy quark part of the electromagnetic current fulfils a Schr\"{o}dinger equation with a potential that has an imaginary part. This imaginary part was bigger than the real part for the situation in which screening is important and hence it indicates that in the perturbative limit the decay width makes the different quarkonium states dissociate before screening becomes relevant.

In the previous discussion we have been talking about potentials and the Schr\"{o}dinger equation without talking about how they can be obtained from QCD. The question can be rephrased in the following way. Is there a controlled expansion in which we can obtain a Schr\"{o}dinger equation as a leading order approximation and we can estimate the error done by making such approximation? A way to answer this question that has been very successful in the vacuum is by using effective field theories (EFTs). This program for finite temperature $T$ and zero chemical potential $\mu$ has been started in the last years \cite{Escobedo:2008sy,Brambilla:2008cx}. Using these techniques the potential of \cite{Laine:2006ns} has been confirmed for $T\gg m_D\sim 1/r$. The case $1/r\gg T\gg E\gg m_D$ was studied in \cite{Brambilla:2010vq} where corrections to decay width that can not be encoded in a potential were found. This result was later compared with lattice computations \cite{Aarts:2011sm} yielding similar results. The case $T\sim 1/r$ was studied in muonic hydrogen \cite{Escobedo:2010tu}. In this work we are going to focus on the recent investigations that have been done to translate the previous results in terms of cross sections \cite{Brambilla:2011sg,Brambilla:2013dpa}. Opposite to what happens to the decay width these cross sections do not depend on the nature of the medium as long as it is isotropic and homogeneous, so they can be directly applied to mediums with a finite chemical potential as the one that is expected to be formed in CBM.

This work is organized as follows. In section \ref{sec:nreft} we review non-relativistic EFTs, in section \ref{sec:gluo} we discuss the process responsible of the decay of quarkonium in the medium at low densities and temperatures that is called gluo-dissociation, in section \ref{sec:ine} we discuss the process that dominates at high densities and temperatures, inelastic parton scattering. Finally, section \ref{sec:concl} is devoted to the conclusions.
\section{Non-relativistic EFTs for heavy quarkonium}
\label{sec:nreft}
The general idea of EFTs was introduced in \cite{Weinberg:1978kz}. An EFT is a field theory which gives exactly the same result as a more general one but on a limited kinematic regime. They are useful for studying problems in which different energy scales are important. In most QCD processes $\Lambda_{QCD}$ is an important scale but in heavy quarkonium we also have another important scale $m_Q$ that fulfils $m_Q\gg \Lambda_{QCD}$. Moreover, because heavy quarkonium is a non-relativistic system, there appears other energy scales as the inverse of the typical radius $1/r\sim m_Q v$ and $E\sim m_Q v^2$ the binding energy. 

Non-relativistic QCD (NRQCD) \cite{Caswell:1985ui} is an EFT in which modes with a virtuality of order $m_Q^2$ are integrated out. The fact that heavy quarks are heavy ensures that NRQCD can be derived from QCD using perturbation theory. In this EFT the heavy quark instead of being represented by a bispinor field is represented by two spinor fields, one for the heavy quark and another for the heavy anti-quark. Approximate symmetries as the spin symmetry can already be seen by looking at the NRQCD lagrangian. 

Potential NRQCD (pNRQCD) \cite{Pineda:1997bj} is an EFT in which modes with a virtuality of order $1/r^2$ are integrated out. In this case pNRQCD can be derived from NRQCD using perturbation theory only for states whose size is smaller than $1/\Lambda_{QCD}$. This is expected to happen for the more deeply bound states as $\Upsilon(1S)$ with a Bohr radius of order $1/a_0\sim 1355 \,\texttt{MeV}$ or $J/\Psi$ with $1/a_0\sim 770\,\texttt{MeV}$ (both values are taken from \cite{pinedathesis}). In the static limit agreement between high orders computations in pNRQCD and the lattice can be found up to distances of order $0.2 \,\textit{fm}$ and in fact it can be used to make a competitive determination of $\alpha_s$ \cite{Bazavov:2012ka}. Another important issue when doing computations is how $\Lambda_{QCD}$ is related with the binding energy $E$. Through this work we are going to assume that $E\gg \Lambda_{QCD}$ when computations at this scale are needed. This might be reasonable for $\Upsilon(1S)$ and $J/\Psi$ where $E\sim 350\,\texttt{MeV}$ \cite{pinedathesis}.

In pNRQCD when $1/r\gg \Lambda_{QCD}$ there exist two kind of fields that represent heavy quark and heavy antiquarks pairs, a singlet field and an octet field. At leading order they both fulfil a Schr\"{o}dinger equation with a potential that encodes physics that is integrated out when going from NRQCD to pNRQCD. 

The main advantage of using EFTs is that one can define a power counting such that the size of each contribution can be predicted just looking at the Lagrangian. As an example let us look in the computation of the binding energy of a given state in perturbation theory. If this is done directly in QCD we will find an expansion of the following type 
\begin{equation}
E=m_Q\alpha_s\sum_{n=0}^\infty\alpha_s^nA_n(v) 
\end{equation}
where $A_n$ are coefficients that in general are going to depend on ratios of the different scale that appear in the problem and therefore on $v$. There is nothing that forbids that a given $A_n$ goes to infinity as $v\to 0$. If this happens we lost control of our perturbative expansion. This is indeed the case of some diagrams that appear in heavy quarkonium physics as for example the box diagram that is enhanced by a $m_Q r$ power. If the same computation is done using pNRQCD the series can be reorganized as
\begin{equation}
E=m_Q\alpha_s v^2\sum_{n,m}\alpha_s^n v^m B_{n,m} 
\end{equation}
where now we know that $B(n,m)$ will be of order $1$. 

With the aim of having a way to correctly define the potential and the Schr\"{o}dinger equation and to have a well defined power counting non-relativistic EFTs were extended to finite temperature. In general the presence of a medium introduces another set of energy scales in the problem and the physics is going to be different depending on the relation with the scales that define the bound state in the vacuum. In a perturbative medium, as the one we are going to consider, the induced scales are the following. There is a scale $p_{Hard}$ which corresponds to the typical energy of the partons in the medium. This energy scale might be integrated out to yield an EFT called Hard Dense Loop (HDL) (Hard Thermal Loop for $\mu=0$) \cite{Braaten:1989mz}. Another important scale is the so-called Debye mass $m_D$ that corresponds to the inverse of the distance in which screening effects become important 
\begin{equation}
m_D^2=\frac{N_c g^2 T^2}{3}+\frac{g^2 T_F N_F}{3}\left(T^2+\frac{3\mu^2}{\pi^2}\right)
\end{equation}

Depending on the relation between $p_{Hard}$ and the rest of scales the dominant mechanism responsible for the decay width is going to be different. It happens that the key relation is that of $m_D$ with the binding energy. If $E\gg m_D$ the dominant mechanism is gluo-dissociation while if $m_D\gg E$ it is inelastic parton scattering. This conclusion can be easily obtained using pNRQCD power counting \cite{Brambilla:2008cx,Escobedo:2010tu}. In the next two sections we are going to discuss these two processes in detail.
\section{Gluo-dissociation}
\label{sec:gluo}
When $E\gg m_D$ the decay width is dominated by the imaginary part of the leading order singlet self-energy. In the case of zero chemical potential this was computed for $T\gg E$ in \cite{Brambilla:2010vq}
\begin{equation}
\delta\Gamma_n=\frac{1}{3}N_C^2C_F\als^3T-\frac{16}{3m}C_F\als TE_n+\frac{4}{3}N_CC_F\als^2T\frac{2}{mn^2a_0} 
\end{equation}
and for $T\sim E$ in \cite{Escobedo:2008sy}
\begin{equation}
\delta\Gamma_n=\frac{4}{3}\als C_FT\langle n|r_i\frac{|E_n-h_o|^3}{e^{\beta |E_n-h_o|}-1}r_i|n\rangle 
\end{equation}

Another way to compute this decay width that has been used in the past is to compute the cross section for this process at $T=0$ and later convolute it with a partonic distribution function. There are two important questions to discuss. If this convolution is justified by perturbative thermal field theory and how the cross section obtained in pNRQCD relates with the one commonly used in phenomenological analyses. The cross section for gluo-dissociation was computed in \cite{Bhanot:1979vb}. They used an OPE based on the multipole expansion. This expansion is also present in pNRQCD Lagrangian so in this sense the result has to be similar. The difference comes from the fact that they use the large $N_c$ limit while this approximation was not needed to compute the decay width in pNRQCD.

In order to extract the cross section from our previous computation of the decay width we use the cutting rules that were generalized to a thermal medium in \cite{Kobes:1986za}. For the case of gluo-dissociation in fact the naively expected convolution is obtained 
\begin{equation}
\Gamma=\int\frac{\,d^3k}{(2\pi)^3}f(k)\sigma(k) \nonumber
\end{equation}
with a cross section for the $1S$ state
\begin{equation}
\sigma_{1S}(k)=\frac{\als C_F}{3} 2^{10} \pi^2  \rho  (\rho +2)^2 \frac{E_1^{4}}{mk^5}
\left(t(k)^2+\rho ^2 \right)\frac{\exp\left(\frac{4 \rho}{t(k)}  \arctan
\left(t(k)\right)\right)}{ e^{\frac{2 \pi  \rho}{t(k)} }-1}\,,
\label{crossEFT}
\end{equation}
where $t(k)\equiv\sqrt{k/\vert E_1\vert-1}$ and $\rho=\frac{1}{N_c^2-1}$. This result was also found in \cite{Brezinski:2011ju} using different techniques. More details about gluo-dissociation in pNRQCD can be found in \cite{Brambilla:2011sg}.
\section{Inelastic parton scattering}
\label{sec:ine}
This process dominates the decay width for $m_D\gg E$. Here we are going to focus in the regime $p_{Hard}\gg 1/r\sim m_D$ that is the more relevant one for dissociation. The process we are studying corresponds to $HQ+p\to Q+\bar{Q}+p$ which means that a heavy quarkonium $HQ$ is hit by a parton $p$ and goes into two heavy quarks $Q$ plus a parton. An approximation that is used in phenomenological analyses \cite{Zhao:2010nk} is $\sigma(HQ+p\to Q+\bar{Q}+p)=2\sigma(Q+p\to Q+p)$ (quasi-free approximation). In the right hand side the cross section computed in \cite{Combridge:1978kx} is used. This approach neglects the interference terms that arise in the interaction of the heavy quark and the heavy antiquark. In the case in which the thermal corrections to the potential can be considered a perturbation one can use the cutting rules to extract a cross section that will not make this approximation. Doing this we notice two things. First that the naive convolution formula can not be applied in this case because we have a parton in the final state
\begin{equation}
\Gamma=\int\frac{\,d^3k}{(2\pi)^3}f(k)(1+f(k))\sigma(k,m_D) 
\end{equation}
and second that $\sigma(k,m_D)$ depends now on the medium through $m_D$ that acts as a thermal mass. For $1S$ states we find 
\begin{equation}
\sigma_{1S}(k,m_D)=8\pi C_F\als^2N_Fa_0^2\left(-\frac{3}{2}+2\log\left(\frac{2}{x}\right)+\log\left(\frac{y^2}{1+y^2}\right)-\frac{1}{y^2}\log(1+y^2)\right)
\end{equation}
where $a_0$ is the Coulombic Bohr radius, $x=m_Da_0$ and $y=ka_0$. In conclusion, the EFT computation takes into account correctly the effects of Pauli blocking and Bose enhancement and improves on the quasi-free approximation. More details about inelastic parton scattering in EFT can be found in \cite{Brambilla:2013dpa}
\section{Conclusions}
\label{sec:concl}
EFTs have been used to compute a wide range of temperature regimes in the case of small chemical potential. This research has shown the importance of the imaginary part of the potential and the decay width for dissociation. In this work we discussed the physical process behind them in different regimes. For gluo-dissociation we found that the pNRQCD improved the previously known cross-section by going beyond the large $N_c$ approximation. For inelastic parton scattering it can be shown that the imaginary part of the potential improves over the quasi-free approximation in taking into account the interference terms.
\section*{Acknowledgements}
I acknowledge Nora Brambilla, Jacopo Ghiglieri and Antonio Vairo for collaboration in the two papers in which this work is based. I also acknowledge financial support from the DFG project BR4058/1-1.

\end{document}